\documentclass[aps,preprint,showpacs, superscriptaddress,onecolumn]{revtex4}%
\usepackage{amsfonts}
\usepackage{amsmath}
\usepackage{amssymb}
\usepackage{graphicx}
\usepackage{subfigure}
\usepackage[svgnames]{xcolor}%
\begin{document}
\preprint{JLAB-THY-07-656}
\preprint{KSUCNR-229-07}
\title{Consequences Of Fully Dressing Quark-Gluon Vertex Function\footnotetext{Notice: Authored by Jefferson Science Associates, LLC under U.S. DOE Contract No. DE-AC05-06OR23177. The U.S. Government retains a non-exclusive, paid-up, irrevocable, world-wide license to publish or reproduce this manuscript for U.S. Government purposes.}\\ With Two-Point
Gluon Lines}
\author{Hrayr H. Matevosyan}
\affiliation{Louisiana State University, Department of Physics \& Astronomy, 202 Nicholson
Hall, Tower Dr., LA 70803, USA}
 \affiliation{Thomas Jefferson National Accelerator Facility, 12000
              Jefferson Ave., Newport News, VA 23606, USA}
\author{Anthony W. Thomas}
\affiliation{Thomas Jefferson National Accelerator Facility, 12000 Jefferson Ave., Newport
News, VA 23606, USA}
\affiliation{College of William and Mary, Williamsburg VA 23187, USA}
\author{Peter C. Tandy}
\affiliation{Center for Nuclear Research, Department of Physics, Kent State University,
Kent, Ohio 44242, USA}
\keywords{$1/N_{c}$ Expansion, $U(1)$ Ghost}
\pacs{12.38.Aw; 11.30.Rd; 12.38.Lg; 12.40.Yx}

\begin{abstract}

We extend recent studies of the effects of quark-gluon vertex dressing  
upon the solutions of the Dyson-Schwinger equation for the quark propagator.  
A momentum delta function is used to represent
the dominant infrared strength of the effective gluon 
propagator so that the resulting integral
equations  become  algebraic.  The guark-gluon vertex is constructed from the complete set of
diagrams involving only 2-point gluon lines.    
The additional diagrams, including those with crossed gluon lines, 
are shown to make an important contribution 
to the DSE solutions for the quark propagator, 
because of their large color factors and the rapid growth in their number.  

\end{abstract}

\date{\today}
\maketitle

\section{Introduction}

In recent years there has been significant progress in the study of the
spectrum of hadrons, as well as their non-perturbative 
structure and form factors,
through approaches that are manifestly covariant and which accommodate both
dynamical chiral symmetry breaking (DCSB) and quark
confinement~\cite{Maris:2003vk}. Covariance provides efficient and unambiguous
access to form factors~\cite{Volmer:2000ek,Maris:2000sk,Alkofer:2004yf} and 
parton distribution functions~\cite{Cloet:2005pp}. 
The covariant approach to modeling QCD with which we will be concerned 
is based upon the Dyson-Schwinger
equations (DSE) of QCD~\cite{Maris:2003vk} .
These are coupled integral equations of motion for
the QCD~\ Green's functions or n-point functions.  
The DSE for any particular n-point 
function involves higher (n+1)-point functions and so on. 
One needs to employ a 
truncation of the coupled equations for practical calculations.

One of the simplest and most studied DSEs is the quark propagator DSE, the
quark gap equation

\begin{equation}
S^{-1}(p)=Z_{2}\,S_{0}^{-1}(p)+{C_{\mathrm{F}}}\,Z_{1}\int_{q}^{\Lambda}
g^{2}D_{\mu\nu}(p-q)\,\gamma_{\mu}
\times S(q)\Gamma_{\nu}(q,p),
\label{gap_eqn}
\end{equation}
where $S_{0}^{-1}(p)=i\gamma\cdot p+m_{bm}$, $m_{bm}$ is the bare current
quark mass, $\Gamma_{\nu}(q,p)$ is the dressed quark-gluon vertex function,
$\int_{q}^{\Lambda}=\int^{\Lambda}d^{4}q/(2\pi)^{4}$ denotes a loop integral
regularized in a translationally-invariant manner at mass-scale $\Lambda$.
Here $Z_{1}(\mu^{2},\Lambda^{2})$ is the vertex renormalization constant to
ensure $\Gamma_{\sigma}=\gamma_{\sigma}$ at the renormalization scale $\mu$ and
$Z_{2}(\mu^{2},\Lambda^{2})$ is the quark wave function renormalization
constant.   The color factor in Eq.~(\ref{gap_eqn}) 
is \mbox{$C_{\mathrm{F}} =
(N_c^2-1)/2N_c$}.

The general form for $S(p)^{-1}$ is
\begin{equation}
S(p)^{-1}=i\gamma\cdot p\,A(p^{2},\mu^{2})+B(p^{2},\mu^{2})
\label{EQ_QUARK_PROPAGATOR}%
\end{equation}
and the renormalization condition at scale $p^{2}=\mu^{2}$ is 
$S(p)^{-1}  
\rightarrow$ $i\gamma\cdot p+m(\mu)$, where $m(\mu)$ is the renormalized
current quark mass.  One can see from Eq.~(\ref{gap_eqn}) that, 
in principle, 
one must first solve the DSE\ for the gluon 2-point function as well as the
DSE for the dressed quark-gluon
vertex.     In the renormalization group improved rainbow 
approximation~\cite{Maris:1997tm},
one replaces the combination of those two objects by bare vertex times an 
effective gluon 2-point function.   
The latter has a phenomenological infrared part
that joins smoothly to the 1-loop renormalization group 
result for the ultraviolet running
coupling.    Typically one infrared parameter is 
sufficient to generate the empirical 
value of the quark condensate.   
When the same kernel and dressed propagators
are used for the Bethe-Salpeter
equation (BSE), this ladder-rainbow (LR) model gives a very good account of
light quark vector and pseudoscalar ground state mesons~\cite{Maris:1999nt}
and many other observables -- see Ref.~\cite{Maris:2003vk} for a review.

As this LR truncation is a non-perturbative truncation, 
one does not have an easy handle 
on its accuracy.    Recent efforts have been made 
to perform calculations  of meson
masses beyond ladder-rainbow truncation~\cite{Bender:2002as,
Bhagwat:2004hn,Matevosyan:2006bk}.    
These studies employ a quark-gluon vertex  that
is dressed by certain classes of diagrams constructed from 2-point 
gluon functions
alone.   The results of a truncation back to a bare 
vertex indicate that the error in the LR
truncation is typically 10-30 percent, 
depending on the class of vertex diagram that 
is taken into account.
To make such calculations feasible, the gluon propagator is simplified
to a momentum delta function so that non-linear 
integral equations are reduced to
non-linear algebraic equations.     The sum of ladder-rainbow diagrams for 
the vertex was implemented in Ref.~\cite{Bender:2002as}, while in 
Ref.~\cite{Bhagwat:2004hn} the important attractive effects of dressing 
with 3-point gluon functions were added phenomenologically. 
The 4-point gluon functions have so far been neglected altogether. 
In our previous work~\cite{Matevosyan:2006bk}, we studied an improved 
ladder-rainbow summation in which internal vertices 
were self-consistently dressed. 
In the current work we explore a scheme
for including all possible diagrams involving vertex dressing with 2-point
gluon functions and explore the effects of this generalization on the behavior
of the quark propagator amplitudes.   
This entails the inclusion of  crossed gluon line diagrams. A study of fully dressed quark-gluon vertex in scalar $\varphi^{2}\chi$ theory with Munczek-Nemirowski gluon propagator of Ref.~\cite{Detmold:2003au} showed that the solutions for quark propagator functions show large variations in the region of the small momenta in comparison with those corresponding to LR truncation of the quark DSE.

\begin{figure}
[ptbh]
\begin{center}
\includegraphics[
height=1.3232in,
width=3.48in
]%
{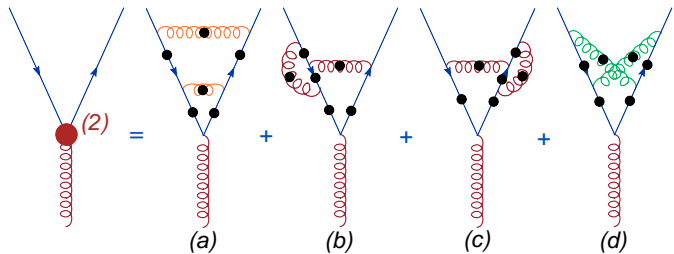}%
\caption{(Color Online) Quark-gluon vertex diagrams 
at $\mathcal{O}(g^{5})$, neglecting
3-gluon and 4-gluon coupling.  
The large red circle denotes the second order vertex and 
the small black circles indicate dressed propagators.  
Diagram~(a) is part of the ladder
sum~\protect\cite{Bender:2002as,Bhagwat:2004hn}, 
while diagrams (b) and (c) are added 
in the improved ladder sum~\protect\cite{Matevosyan:2006bk}, 
and diagram~(d) enters 
in the extension considered in this work.   
Crossed gluon diagrams, such as (d), prove to 
be important because of their significantly larger color factors.}%
\label{PLOT_FV_ORDER_2}%
\end{center}
\end{figure}

To demonstrate the relevance of the current dressing scheme, consider the
vertex with exactly $2$ gluon lines, Fig. \ref{PLOT_FV_ORDER_2}. In our
previous work with the most general vertex dressing scheme to 
date~\cite{Matevosyan:2006bk}, we extended the ladder-summed approximation 
of Refs.~\cite{Bender:2002as,Bhagwat:2004hn} (Fig.~\ref{PLOT_FV_ORDER_2}a) 
to self-consistently include the dressing of internal vertices 
(Figs.~\ref{PLOT_FV_ORDER_2}b and \ref{PLOT_FV_ORDER_2}c).  The same procedure 
was extended to a general number of gluon lines 
and the contributions were summed 
until convergence was  
reached for this improved ladder-rainbow sum.   Crossed gluon line diagrams, 
such as Fig.~\ref{PLOT_FV_ORDER_2}d were not included.     Convergence 
in this improved ladder-rainbow sum was found in the vertex itself, 
in the quark 
DSE solutions, and in the pseudoscalar and vector meson masses.  

The diagrams in the improved ladder-rainbow sum have the same simple color
structure and the associated color factors are powers of one basic color factor.
The color structure analysis of the diagrams 
omitted from the improved ladder-rainbow
class shows that they can have significantly 
larger color-factors and that, at the very least, their role should 
be explored. 
For example, with $N_{c}$ colors, the color factor associated 
with Figs.~\ref{PLOT_FV_ORDER_2}a, 
\ref{PLOT_FV_ORDER_2}b and \ref{PLOT_FV_ORDER_2}c are
$1/(2N_{c})^{2}$, whereas for the crossed gluon 
diagram of Fig.~\ref{PLOT_FV_ORDER_2}d 
the color factor is $(N_{c}^{2}+1)/(2N_{c})^{2}$.  
In general, similar examples occur at 
higher order.    Also working to magnify the contributions to the vertex from 
many gluon lines is the fact that the number of such diagrams increases rapidly. 
A formal analysis of these additional diagrams in terms of  $1/N_{c}$ expansion
is presented in Ref.~\cite{Matevosyan:2007wc}.

\section{The Fully Dressed Vertex With Two-Point Gluon Lines}

We employ Landau gauge and a Euclidean metric, with: $\{\gamma_{\mu}%
,\gamma_{\nu}\}=2\delta_{\mu\nu}$; $\gamma_{\mu}^{\dagger}=\gamma_{\mu}$; and
$a\cdot b=\sum_{i=1}^{4}a_{i}b_{i}$. The dressed quark-gluon vertex for gluon
momentum $k$ and quark momentum $p$ can be written $ig\,t^{c}\,\Gamma_{\sigma
}(p+k,p)$, where $t^{c}=\lambda^{c}/2$ and $\lambda^{c}$ is an SU(3) color
matrix. In general, $\Gamma_{\sigma}(p+k,p)$ has 12 independent invariant
amplitudes. We are particularly concerned in this work with the vertex at
$k=0$, in which case the general form is%
\begin{equation}
\Gamma_{\sigma}(p)=\alpha_{1}(p^{2})\gamma_{\sigma}+\alpha_{2}(p^{2}%
)\gamma\cdot p~p_{\sigma}-\alpha_{3}(p^{2})ip_{\sigma}\label{vertex_form}
+\alpha_{4}(p^{2})i\gamma_{\sigma}~\gamma\cdot p
\end{equation}
where $\alpha_{i}(p^{2})$ are invariant amplitudes.

We propose a scheme for dressing the vertex function that includes all
possible contributions constructed with two-point 
gluon lines up to $n$ 
gluon lines. We can decompose the vertex function 
into a sum of all contributions
with exactly $i$ gluon lines:%
\begin{equation}
\Gamma_{\mu}=\sum_{i=0}\,\Gamma_{\mu}^{i}, \label{EQ_VERT_DECOMP}%
\end{equation}
where $\Gamma_{\mu}^{0}=Z_{1}\,\gamma_{\mu}$.

In previous work we developed a scheme to include all 
vertex diagrams that have a ladder structure 
but further improved it by the self-consistent dressing of all internal 
vertices~\cite{Matevosyan:2006bk}.   
In that case the contribution $\Gamma_{\mu}^{i}$
is generated from three vertex contributions having a
smaller number of gluon lines by adding one gluon ladder rung with dressed
vertices. If the number of gluon lines in the three vertex contributions are
denoted $j,k$ and $l$, then summation is made over $j,k$ and $l$ such that
$j+k+l+1=i$.  That iterative self-consistent ladder scheme is described by
\begin{multline}
\Gamma_{\mu}^{i}(p+k,p)=-\left(-\frac{1}{2N_c}\right) \sum
_{\substack{j,k,l\\i=j+k+l+1}}\int_{q}^{\Lambda}g^{2}D_{\sigma\nu
}(p-q)\label{EQ_PLANAR_VERTEX_ITERATION}\\
\times\Gamma_{\sigma}^{j}(p+k,q+k)S(q+k)\Gamma_{\mu}^{l}(q+k,q)
S(q)\Gamma_{\nu}^{k}(q,p),
\end{multline}
for $i\geq1$.    Here, and in all vertex constructions that we 
discuss and implement, the
quark propagators that are internal to the vertex 
are self-consistently dressed  using the gap equation,
Eq.~(\ref{gap_eqn}), with the dressed vertex at the order being discussed.  
The color factors of all contributions included in this improved ladder vertex 
are correctly generated by the iterations of 
\mbox{$-\frac{1}{2N_c}$} implied by the above.
Such a vertex, generated by 2-point gluon lines, 
is repulsive and as a consequence one 
finds~\cite{Bhagwat:2004hn} that the dominant vertex amplitude 
\mbox{$\alpha_{1}(p^{2}) < 1$}.    

However, the available lattice data indicate that 
\mbox{$\alpha_{1}(p^{2}) > 1$} much as 
it is in an Abelian theory.    Certainly in the 
ultraviolet, where a one-loop analysis is 
reliable~\cite{Bhagwat:2004kj},
$\alpha_{1}(p^{2}) \to 1$ from above because of 
the strong attraction is provided by the  3-gluon 
coupling term with its dominant and large color factor $+\frac{N_c}{2}$.   
In Refs.~\cite{Bhagwat:2004hn,Matevosyan:2006bk},  
it was argued that an effective
way to simulate such physics in studies like the present 
is to assume the momentum 
dependence of the corresponding 3-gluon coupling contribution 
to the vertex is similar enough to 
Eq.~(\ref{EQ_PLANAR_VERTEX_ITERATION}) to allow the color factor 
there to be replaced
by the sum of the two, which is $C_{\mathrm{F}}$, 
the color  factor of the quark DSE in 
Eq.~(\ref{gap_eqn}).  To reproduce lattice vertex data, 
and to explore the effect of 
3-gluon coupling in a slightly more general way, 
Refs.~\cite{Bhagwat:2004hn,Matevosyan:2006bk},  
employed Eq.~(\ref{EQ_PLANAR_VERTEX_ITERATION}) with color factor 
$\mathcal{C}{C_{\mathrm{F}}}$ where $\mathcal{C}$ varies in the 
range $-\frac{1}{2N_c \;{C_{\mathrm{F}}} }$ to 1.    Here we will initially use the lower limit which corresponds to omitting all consideration
of 3-gluon coupling.   Subsequently, with the choice 
\mbox{$\mathcal{C} = 0.5$}, we explore the
likely influence of 3-gluon coupling with the present vertex.   

In this present work we explore the addition of the 
remaining class of vertex diagrams that
are beyond those in the improved ladder sum outlined above, 
but still obtainable in terms of  2-point gluon functions.   

\begin{figure}
[ptb]
\begin{center}
\includegraphics[
height=1.7417in,
width=0.9496in
]%
{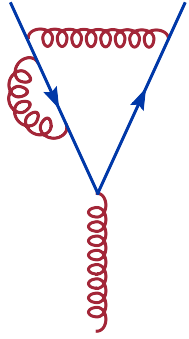}%
\caption{(Color Online) Quark self-energy type contribution in dressing the
quark-gluon vertex.}%
\label{PLOT_FV_SELFEN}%
\end{center}
\end{figure}

\begin{figure}
[ptbptb]
\begin{center}
\includegraphics[
height=4.9286in,
width=4.7297in
]%
{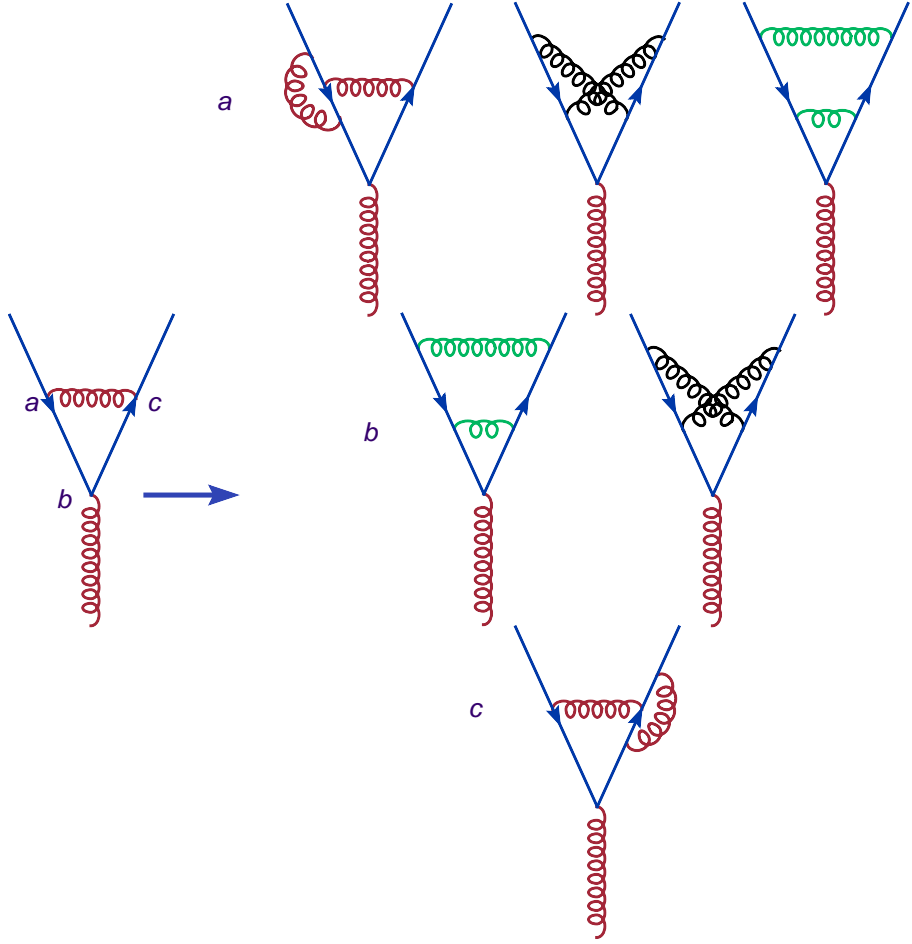}%
\caption{(Color Online) Generating all the second order diagrams from the only
first order diagram by making all possible insertions of a single gluon line
into the first order vertex diagram. The letters indicate the starting point
of the inserted line. We note that there are several redundant diagrams
generated, which must be removed.}%
\label{PLOT_FV_2GEN}%
\end{center}
\end{figure}

To generate such a vertex, we use a recursive algorithm for 
obtaining all the contributions with exactly
$i$ gluon lines, $\Gamma_{\mu}^{i}$ , from the diagrams contributing to
$\Gamma_{\mu}^{i-1}$.  We take each diagram contributing to $\Gamma_{\mu
}^{i-1}$ and consider the diagrams produced by making all possible insertions
of a single gluon line so that it starts and ends on a fermion 
line without being simply 
a fermion self-energy insertion (see Fig. (\ref{PLOT_FV_SELFEN})). 
Then, to avoid double 
counting, a check is made 
as to whether the resulting diagram was already produced from 
any previous construction 
at the same order. One can easily
see that the algorithm works for constructing the one-loop diagram from the
bare vertex and from that the two-loop diagrams 
of Fig.~(\ref{PLOT_FV_ORDER_2}).  
The steps of the algorithm are illustrated in Fig. (\ref{PLOT_FV_2GEN}).%

It is trivial to prove, using mathematical induction, that this algorithm
produces \textit{all} possible diagrams involving $n$ gluon lines in dressing
the vertex for any $n\geq1$. In the previous paragraph we demonstrated that
all the first and second order diagrams are generated using this algorithm.
Let us suppose, for some $n>1$, that we start with the 
correct diagrammatic content of
$\Gamma_{\mu}^{n-1}$ but that the algorithm for $\Gamma_{\mu}^{n}$ failed to 
produce a certain valid diagram with $n$ gluon lines.  It
is easy to see that we can always find one gluon line, which can be removed
from that diagram so that the remaining $n-1$ gluon lines are not simply quark 
self-energy insertions.  Hence the produced parent vertex is a valid member of  
$\Gamma_{\mu}^{n-1}$ and the supposition that all possible 
insertions into it are not
capable of generating  all members of $\Gamma_{\mu}^{n}$ is incorrect.

In fact, the gluon line that can be removed is easy to find in a diagram. If
a gluon line attached to the end of a quark line cannot
be so removed, then the other end of this gluon line separates ends of a second gluon
line that would otherwise be a self-energy insertion.  This second gluon line is the
one that can be removed.

\begin{figure}
[ptb]
\begin{center}
\includegraphics[
height=1.7262in,
width=2.1188in
]%
{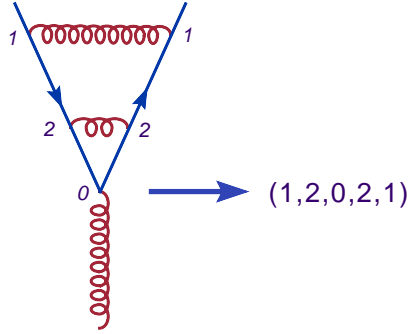}%
\caption{(Color Online) Constructing the set of numbers uniquely identifying
one of the second order diagrams.}%
\label{PLOT_FV_2ORD_UNSET}%
\end{center}
\end{figure}

The algorithm described above is easy to implement in practice by constructing
a set of numbers uniquely identifying each diagram. We build the set by
enumerating the bare quark-gluon vertices in a diagram with $n-1$ gluon lines
from $1$ to $n-1$ and assigning the same numbers to the vertices attached to
the same gluon propagators. We assign $0$ to the external gluon vertex. An
example of such construction is shown in Fig.~(\ref{PLOT_FV_2ORD_UNSET}).

\begin{figure}
[ptb]
\begin{center}
\includegraphics[
height=0.7057in,
width=3.5829in
]%
{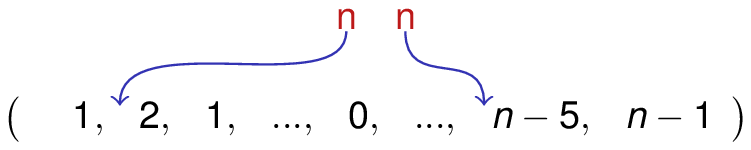}%
\caption{(Color Online) The sets representing all  
diagrams with exactly $n$ gluon lines are
obtained from those corresponding to diagrams with exactly $\ n-1$ lines by
inserting a pair of $n$'s into the set and making all possible permutations, so
they are never next to each other. }%
\label{PLOT_FV_SET_INSERT}%
\end{center}
\end{figure}

To construct the vertices with $n$ gluon lines 
we insert a pair of integers $n$ into
the set described above, so that they are not next to each other, as
illustrated in Fig.~(\ref{PLOT_FV_SET_INSERT}). We relabel the resulting
set in ascending order and check whether the final set was already
generated.

\subsubsection{Algebraic Analysis}

In order to solve the gap\ equation with our dressed quark-gluon vertex one
needs the gluon 2-point function. As in the previous studies of this kind, we
again employ the Munczek-Nemirovsky Ansatz~\cite{Munczek:1983dx} for the
interaction kernel in the form 
\begin{equation}
g^{2}D_{\mu\nu}(k)\rightarrow\left(  \delta_{\mu\nu}-\frac{k_{\mu}k_{\nu}%
}{k^{2}}\right)  (2\pi)^{4}\mathcal{G}^{2}\delta^{4}(k),
\label{EQ_GLUON_PROPAGATOR}%
\end{equation}
where the parameter $\mathcal{G}^{2}$ is a measure of the integrated kernel strength.
This simplification reduces the multi-dimensional integral equations for the vertex and 
the dressed quark propagator to algebraic ones that make the analysis feasible.

The algebraic form of the gap equation for the quark propagator corresponding
to the gluon propagator in Eq.~(\ref{EQ_GLUON_PROPAGATOR}) is
\begin{equation}
S^{-1}(p)=i\gamma\cdot p+m_{bm}+\mathcal{G}^{2}\gamma_{\mu}S(p)\Gamma_{\mu
}(p). 
\label{EQ_gap_EQUATION_SIMPL}
\end{equation}

After projection onto the two Dirac amplitudes we have
\begin{align}
A(p^{2})  &  =1-\mathcal{G}^{2}\frac{i}{4}tr\left[  \frac{\gamma\cdot p}%
{p^{2}}\gamma_{\mu}S(p)\,\Gamma_{\mu}(p)\right]  ,\label{EQ_PROPAGATOR_A}\\
B(p^{2})  &  =m_{bm}+\mathcal{G}^{2}\frac{1}{4}tr\left[  \gamma_{\mu
}\,S(p)\,\Gamma_{\mu}(p)\right]  . \label{EQ_PROPAGATOR_B}%
\end{align}
Equations~(\ref{EQ_PROPAGATOR_A}) and (\ref{EQ_PROPAGATOR_B}) are solved
simultaneously with a vertex function calculated at a specified order $n$ of
vertex dressing.

The calculations of the Dirac algebra for 
the vertex diagrams was performed using
computer-algebraic methods provided by the \textit{FeynCalc} package
\cite{Mertig:1990an} for \textit{Mathematica }\cite{Wolfram}. The color
factors were calculated by numerical contraction of 
the SU(3) color matrices. The number
of possible diagrams grows extremely fast with the number of gluon lines
included, making it difficult to advance too far with the number of loops. For
example, while there are only $4$ possible diagrams with exactly $2$ internal
gluon lines, there are $27$ with $3$ and $38,232$ with $6$ internal lines. Such
a rapid increase in the number of possible contributions forced us to use
parallel computing to calculate the invariant amplitude functions for the
vertex with exactly $6$ gluon lines within a reasonable time frame. The
contributions to $\Gamma_{\mu}^{7}$ include more than $5\times10^{5}$ possible
diagrams, which we did not consider feasible to calculate. The comparison of
the number of diagrams with the previous dressing schemes is shown in 
Table~\ref{TBL_FV_COM}.

\begin{table}[ptb]
\caption{Comparison of the number of diagrams with exactly $n$ gluon lines
included in various vertex dressing schemes:  ladder rainbow 
summed~\protect\cite{Bender:2002as,Bhagwat:2004hn},  self-consistently improved ladder 
rainbow summed~\protect\cite{Matevosyan:2006bk}, and  the present full vertex obtainable with
just 2-point gluon lines.}%
\label{TBL_FV_COM}%
$\left(
\begin{array}
[c]{c|c|c|c}%
n & \textcolor{SeaGreen}{LR \hspace{1 pt} Summed} &
\textcolor{SteelBlue}{ Improved} & \textcolor{Maroon}{Full}\\\hline
2 & \textcolor{SeaGreen}{1} & \textcolor{SteelBlue}{ 3} &
\textcolor{Maroon}{4}\\\hline
3 & \textcolor{SeaGreen}{1} & \textcolor{SteelBlue}{ 6} &
\textcolor{Maroon}{27}\\\hline
6 & \textcolor{SeaGreen}{1} & \textcolor{SteelBlue}{ 21} &
\textcolor{Maroon}{38,232}\\\hline
7 & \textcolor{SeaGreen}{1} & \textcolor{SteelBlue}{ 28} &
\textcolor{Maroon}{\sim5*10^5}
\end{array}
\right)  $\end{table}

\begin{figure}[ptb]
\begin{center}
\includegraphics[width=0.8\textwidth
]{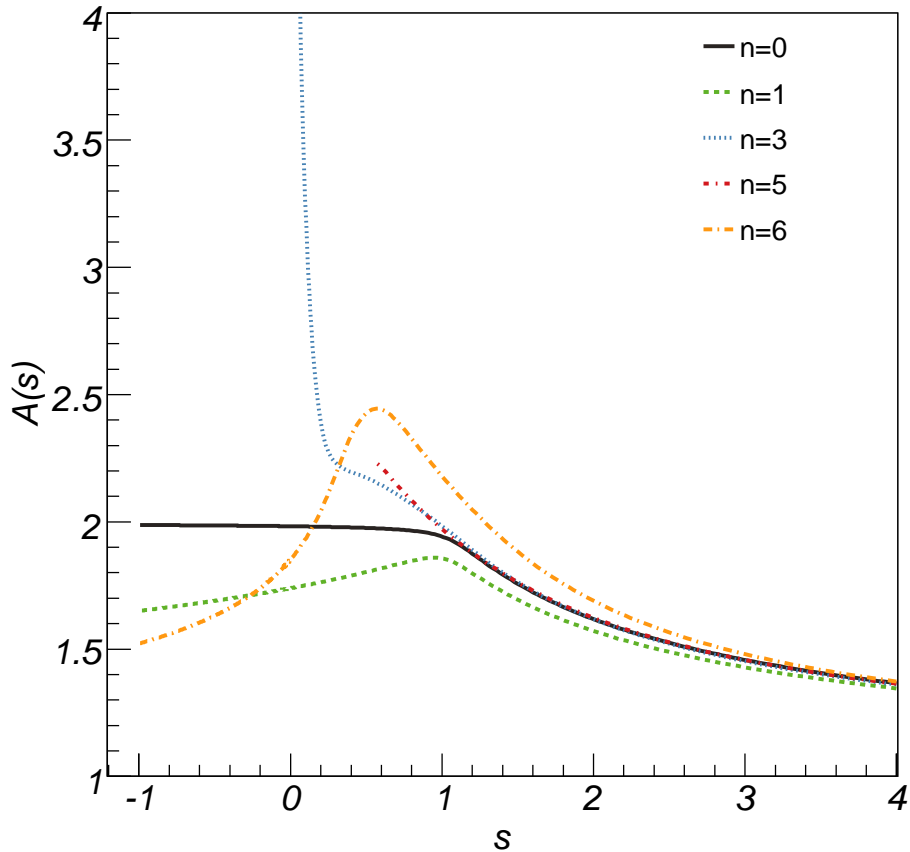}
\end{center}
\caption{(Color Online) Quark propagator amplitude $A(s)$ versus Euclidean
$s=p^{2}$. We use the interaction mass scale $\mathcal{G}=1~\operatorname{GeV}%
$ and the current mass is $m=0.0175~\mathcal{G}=17.5\,$~$\operatorname{MeV}$.
We show the influence of vertex dressing 
to order $n$ in the number of gluon lines
as described in the
text. $n=0$ yields the solid curve and the result is the ladder-rainbow
truncation. The other curves are $n=1$ (long dashed curve, 1-loop vertex),
$n=3$ (short dashed curve), $n=5$ (dot short dashed curve) and $n=6$ (dot long
dashed curve).}%
\label{PLOT_A_DIFF_ORD}%
\end{figure}

\begin{figure}[ptbptb]
\begin{center}
\includegraphics[width=0.8\textwidth
]{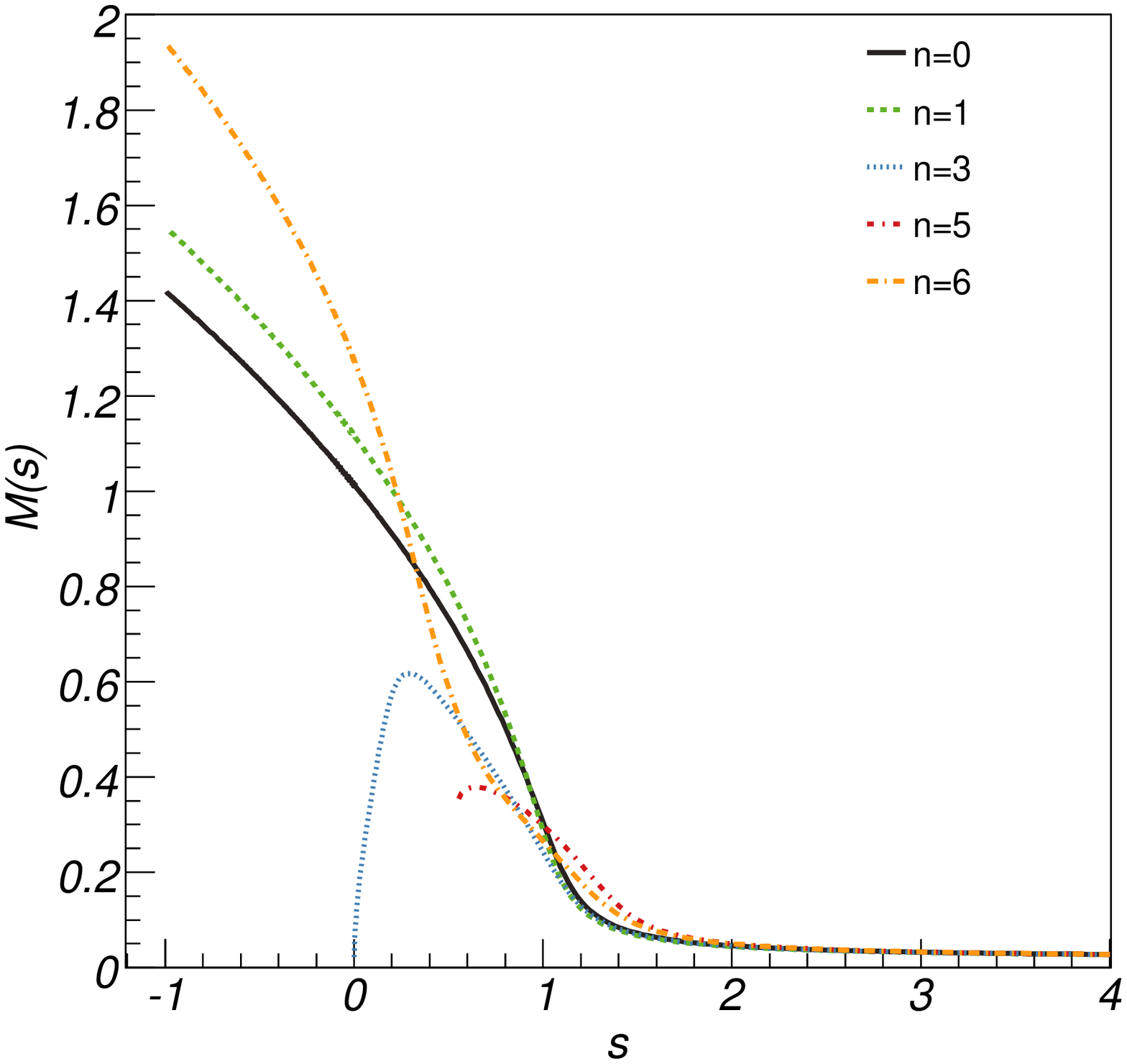}
\end{center}
\caption{(Color Online) Quark mass function $M(s)$ versus Euclidean $s=p^{2}$.
We use the interaction mass scale $\mathcal{G}=1~\operatorname{GeV}$ and the
current mass is $m=0.0175~\mathcal{G}=17.5\,$~$\operatorname{MeV}$. We show
the influence of vertex dressing to order $n$ in the number of gluon lines 
as described in the text.  $n=0$ yields the solid curve and the result is 
the ladder-rainbow truncation. The
other curves are $n=1$ (long dashed curve, 1-loop vertex), $n=3$ (short dashed
curve), $n=5$ (dot short dashed curve) and $n=6$ (dot long dashed curve).}%
\label{PLOT_M_DIFF_ORD}%
\end{figure}

The numerical solutions of the gap equation constructed with the full vertex
are shown in Figs.~\ref{PLOT_A_DIFF_ORD} and \ref{PLOT_M_DIFF_ORD}
for a current quark mass \mbox{$m = $}~ 17 MeV.       No attempt has been made
to simulate 3-gluon coupling -- i.e., the color factors 
used are as directly given by 
the $SU(3)_c$ algebra.     
As found in previous analysis with this model, the results for even and odd $n$
behave differently.   The even orders provide a behavior for $A(s)$ and
$M(s)$ that is analytic and well-defined in the 
infrared spacelike region continuing
into the timelike region \mbox{$s> -1$} explored here.     
Mass generation is strong
and investigations of the approach to the 
chiral limit \mbox{$m \to 0$} indicate
that these solutions are continuously connected to the Nambu-Goldstone mode
of spontaneous breaking of chiral symmetry.   

The odd $n$ solutions \mbox{$n \geq 3$} develop non-analytic 
points in the infrared.     
There is also a tendency for significantly weaker 
mass generation in such solutions.    
Evidently, repulsive color factors dominate for odd $n$ 
more than for even $n$
and the mass generation effects are in accord with that.      

There is no convincing evidence of convergence 
with respect to increasing $n$.   
At least two factors are at work here.   Firstly, with 
the increasing number $n$ of gluon lines, 
there are diagrams at each order that have 
color factors that are comparable to 
that of the  \mbox{$n = 1$} contribution to the vertex.    
The class of diagrams 
omitted in previous studies of this type, e.g., Fig.~\ref{PLOT_FV_ORDER_2}d, 
contains members that have larger color factors than 
the improved ladder-rainbow
class.   Secondly, the number of  diagrams specified 
by $n$  increases rapidly with 
$n$, as illustrated in Table \ref{TBL_FV_COM}.


\subsubsection{Effect of 3-point gluon function dressing}

Explicit calculations with gluon 3-point functions proved to be extremely
difficult, both from the point of view of reproducing all possible diagrams at
some non-trivial order in quark-gluon coupling and from the fact that no
suitable, model gluon 2-point function is known to us which would allow for
feasible calculations beyond one or two loops. Hence we resort to the method
introduced in Ref.~\cite{Bhagwat:2004hn}, which allows 
one to effectively  account for 
3-point gluon function dressing of the vertex through the introduction of the
phenomenological parameter $\mathcal{C}$, which for \mbox{$N_c = 3$} can 
take values in the range $-1/8\leq\mathcal{C}\leq1$, as discussed  after 
Eq.~(\ref{EQ_PLANAR_VERTEX_ITERATION}).    

\begin{figure}[ptb]
\begin{center}
\includegraphics[width=0.8\textwidth
]{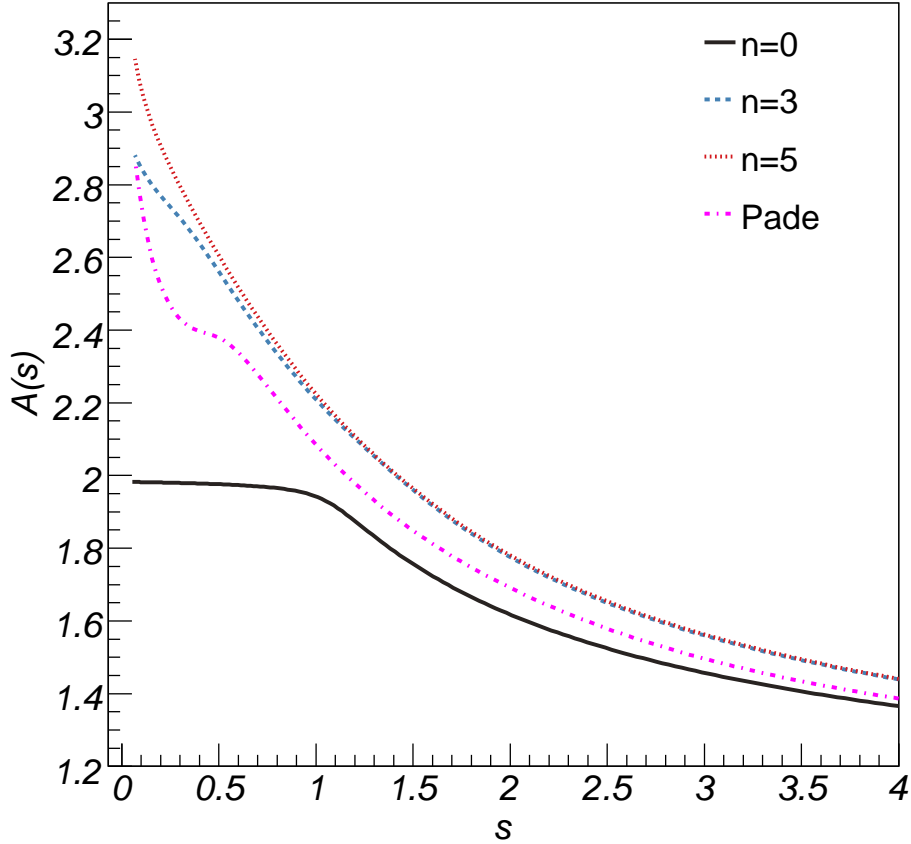}
\end{center}
\caption{(Color Online) Quark propagator amplitude $A(s)$ versus Euclidean
$s=p^{2}$. We use the interaction mass scale $\mathcal{G}=1~\operatorname{GeV}%
$, the current mass is $m=0.0175~\mathcal{G}=17.5\,$~$\operatorname{MeV}$.  
Partial account has been taken of 3-gluon coupling through the parameter
$\mathcal{C}=0.5$ which induces more mass generation.   The influence of 
quark-gluon vertex dressing to order $n$ in the number of gluon lines:  
$n=0$ (solid curve, ladder
rainbow truncation),  $n=3$ (long dashed curve), and $n=5$ (short dashed curve).
The dot dashed curve shows the Pad\'{e} approximation to the vertex. }%
\label{PLOT_A_EC}%
\end{figure}

\begin{figure}[ptbptb]
\begin{center}
\includegraphics[width=0.8\textwidth
]{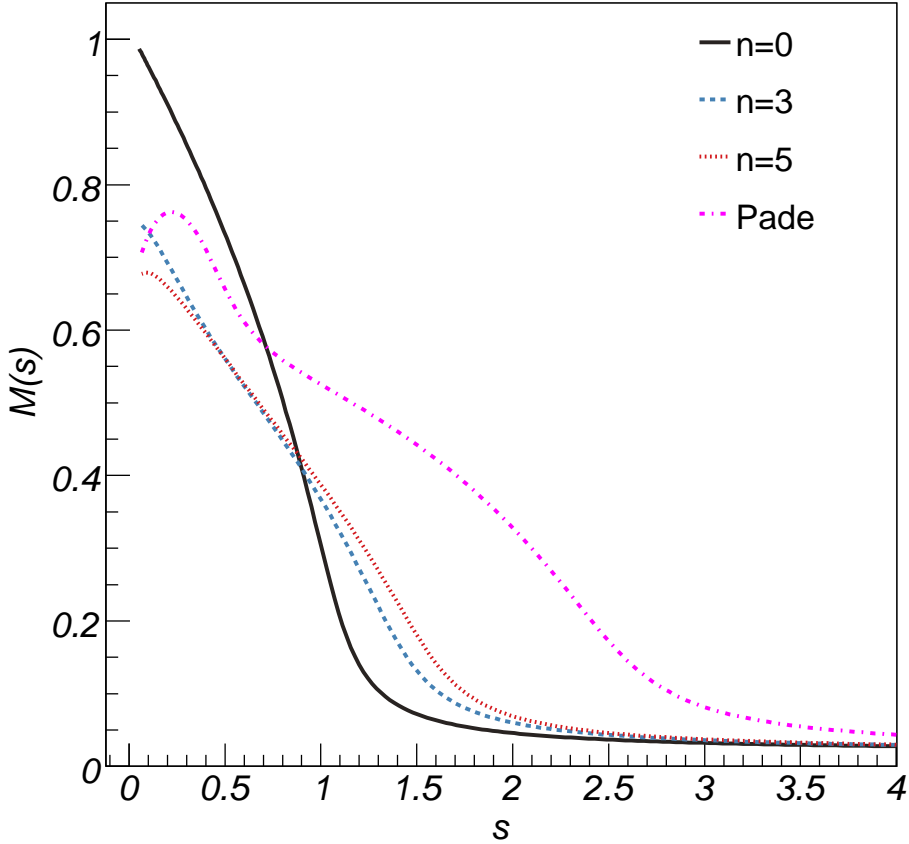}
\end{center}
\caption{(Color Online) Quark propagator amplitude $M(s)$ versus Euclidean
$s=p^{2}$. We use the interaction mass scale $\mathcal{G}=1~\operatorname{GeV}%
$, the current mass is $m=0.0175~\mathcal{G}=17.5\,$~$\operatorname{MeV}$.  
Partial account has been taken of 3-gluon coupling through the parameter
$\mathcal{C}=0.5$ which induces more mass generation.   The influence of 
quark-gluon vertex dressing to order $n$ in the number of gluon lines:  
$n=0$ (solid curve, ladder
rainbow truncation),  $n=3$ (long dashed curve), and $n=5$ (short dashed curve).
The dot dashed curve shows the Pad\'{e} approximation to the vertex. }%
\label{PLOT_M_EC}%
\end{figure}

We chose to implement this phenomenological scheme only for the sub-class of 
vertex diagrams that correspond to the improved ladder structure where its 
motivation is unambiguous.    We note that all such diagrams have the
same color factors as illustrated in Eq.~(\ref{EQ_PLANAR_VERTEX_ITERATION}).
The results show that for the parameter range 
$\mathcal{C}\in(0.375,$ $0.8)$ the solutions of 
the gap equation have significant 
mass generation even for the odd $n$ values.    
This is indicative of the attractive
effect of the 3-gluon coupling mechanism.   Also, in this circumstance there
is a much weaker tendency of the odd orders \mbox{$n \geq 3$} to develop 
non-analytic points in the deep infrared.     The solutions for
$\mathcal{C}=0.5$ are presented 
in Figs.~\ref{PLOT_A_EC} and \ref{PLOT_M_EC}.%



\subsubsection{Pad\'{e} Approximant}

One can see from Figs.~(\ref{PLOT_A_EC}) and (\ref{PLOT_M_EC}) that even
with the enhanced mass generation generated by the strong attractive role
of the effective 3-point gluon coupling, the solutions do not show 
convincing convergence with respect to the maximum number, $n$, of gluon lines
allowed in the quark-gluon vertex dressing, even in the space-like region. 
To seek more reliable information, we employ a Pad\'{e} approximant to
re-sum the quark self-energy contributions with respect to $n$.  
While the details of this calculation are presented in 
Appendix A, the graphs corresponding to the solutions of the gap\ equation
with Pad\'{e} approximant used for the quark self-energy contribution are
shown in magenta in Figs.~(\ref{PLOT_A_EC}) and (\ref{PLOT_M_EC}). It is clear
from the graph that our results are significantly different from those 
obtained using LR\ truncation and illustrate that convergence is still remote 
for \mbox{$n \approx 5$}.   

Unfortunately the present model does not allow consistent solutions in the time-like region, where the propagators are needed in order to calculate the meson masses by solving the
corresponding Bethe-Salpeter equations.

\section{Summary}

We have extended recent studies of the effects of quark-gluon vertex dressing  
upon the solutions of the Dyson-Schwinger equation for the quark propagator.  
The approach we have followed uses a momentum delta function to represent
the dominant infrared strength of the effective gluon 
propagator so that the resulting integral
equations  become  algebraic.  When the guark-gluon vertex is constructed from
diagrams that involve only 2-point gluon lines, the algebraic form allows the
summation of all vertex diagrams, self-consistently with solution
of the quark propagator.    Previous studies along these lines have treated
the ladder-rainbow class of vertex diagram, and subsequently an improved
ladder-rainbow class in which all internal vertices are self-consistently 
dressed.   Whereas the earlier improvements showed only minor differences from
the Ladder-Rainbow approximation, the inclusion of all two gluon lines at a given 
order leads to much larger corrections in the infrared space-like as well as the time-like
region. 

As far as we are able to take 
the calculations, there is no evident
convergence with increasing $n$. In particular,
the behavior with $n$ even or odd is rather different.  
At least two factors are at work here.   Firstly, with 
increasing $n$, 
there are diagrams at each order that have color factors 
which are comparable  with 
that of the  \mbox{$n = 1$} contribution to the vertex.    
The class of diagrams 
omitted in previous studies of this type, 
contains members that have larger color factors 
than the improved ladder-rainbow
class of diagrams.   Secondly, the number of  diagrams that must be included   
increases rapidly with $n$.

The resulting solutions of the gap equation with the corresponding
vertex calculated up to a  given order, $n$, in gluon lines show 
significant deviation from
those calculated previously using either a 
bare vertex or a vertex dressed according
to a ladder-rainbow sum or its self-consistent improvement. The investigation
of the infrared behavior of the ghost and gluon n-point functions of Refs. \cite{Alkofer:2004it, Alkofer:2006gz} indicate that the contributions of ghost loops in dressing the quark-gluon vertex are dominant.
It is possible  that a
consistent treatment involving 2-, 3-, and 4-point gluon functions  as well as ghost contributions might enable
a convergent summation of diagrams for the vertex.      However the present 
simplified gluon propagator permits only zero momentum and this is not
compatible with an explicit treatment of ghost 2-point, and 3- and 4-point 
gluon functions.   In spite of the the numerous interesting applications 
of the Munczek-Nemirovsky model, it is also
possible that the advantages of the simple gluon propagator, 
in allowing algebraic
simplification of the summation of vertex diagrams within the quark DSE , are 
outweighed by  model artifacts. It would be very important to explore ways
to repeat this calculation with a more realistic gluon propagator.     

\begin{acknowledgments}
This work was supported in part by DOE contract DE-AC05-06OR23177, 
under which Jefferson Science Associates, LLC,
operates Jefferson Lab, the U.S. National Science Foundation under Grant
Nos. PHY-0500291 and PHY-0610129 and the Southeastern Universities Research
Association(SURA). HHM thanks Jerry P. Draayer for his support during the
course of the work, the Graduate School of Louisiana State University for a
fellowship partially supporting his research, George S. Pogosyan and Sergue I.
Vinitsky for their support at the Joint Institute of Nuclear Research. 
\end{acknowledgments}

\appendix

\section{Pad\'{e} Approximant for the quark self-energy\label{APD_3A}}

The solutions of the gap\ equation (\ref{EQ_gap_EQUATION_SIMPL}) for the quark
propagator amplitudes $A(s)$ and $B(s)$ depend on the number $n$ of the
included gluon lines in the calculations through the Dirac projections of the
quark self-energy term with appropriately dressed quark-gluon vertex function
on the RHS of Eqs.(\ref{EQ_PROPAGATOR_A}) and (\ref{EQ_PROPAGATOR_B}) (the
terms under\ the tr). We re-summed the perturbative solutions of the
gap\ equation and obtained a\ solution at $n=\infty$ by employing a Pad\'{e}
approximant in $n$ for these self-energy terms.

In order to construct a Pad\'{e} approximant for each of these terms we
established their dependence on $n$ by introducing a fictitious coefficient,
$\omega$, to the bare quark-gluon vertex coupling $g$ in dressing the vertex
function and constructing the projections of the self-energy term in Eqs.
(\ref{EQ_PROPAGATOR_A}) and (\ref{EQ_PROPAGATOR_B}). In each case a Pad\'{e}
approximant of the following form%
\begin{equation}
f(\lambda)=\frac{a_{0}+a_{1}\lambda+a_{2}\lambda^{2}}{1+b_{1}\lambda
+b_{2}\lambda^{2}}, \label{EQ_PADE_FORM}%
\end{equation}
where $\lambda=\omega^{2}$, was used to match the appropriate self-energy
projection term by solving for the coefficients $a_{i}$ and $b_{i}$. We
considered the self-energy projection terms with the quark-gluon vertex
dressed to order $n=3$ in the number of gluon loops to determine all the
parameters in (\ref{EQ_PADE_FORM}).

After determining the coefficients $a_{i}$ and $b_{i}$,\ the parameter
$\omega$ was set to $1$ and the corresponding coupled equations for $A(s)$ and
$B(s)$ were solved. For the case $\mathcal{C}=0.5$, the solutions obtained
with the approximant self-energy projection terms are finite and continuous in
the space-like region.

\bibliographystyle{apsrev}
\bibliography{DSE-LN}

\end{document}